\documentclass{IEEEtran}
\usepackage{amssymb}
\usepackage{}
\usepackage{amsfonts}
\usepackage{amsmath}
\usepackage{algorithm}
\usepackage{cite}

\usepackage{graphicx,subfigure,amsmath,amssymb,cite}

\usepackage[dvips]{color}
\usepackage{float}
\ifCLASSINFOpdf
\else
\fi
\begin{document}
\title{Secure Directional Modulation to Enhance Physical Layer Security in IoT Networks}

\author{Feng~Shu,~Siming~Wan,~Shihao Yan, Qian~Wang,~Yongpeng~Wu,~Riqing~Chen,~Jun~Li, ~and~Jinhui~Lu

\thanks{Feng Shu,~Siming Wan,  Jun Li, and Jinhui Lu are with School of Electronic and Optical Engineering, Nanjing University of Science and Technology, Nanjing, 210094, China (emails: \{shufeng, wsm, jun.li, jinhui.lu\}@njust.edu.cn).}
\thanks{Feng Shu and Riqing~Chen are also with the College of Computer and Information Sciences, Fujian Agriculture and Forestry University, Fuzhou 350002, China, and the College of Physics and Information, Fuzhou University, Fuzhou 350116, ~China (email: chenriqing1980@163.com).}
\thanks{Shihao Yan is with the Research School of Engineering, The Australian National University, Canberra, Australia (email: shihao.yan@anu.edu.au).}
\thanks{Qian Wang is with  the School of Computer Science, Wuhan University, Wuhan, China (email: qianwang@whu.edu.cn).}
\thanks{Yongpeng Wu is with the Shanghai Key Laboratory of Navigation and Location-Based Services, Shanghai Jiao Tong University, Minhang 200240, China (email: yongpeng.wu2016@gmail.com).}}

\maketitle

\begin{abstract}
In this work, an adaptive and robust null-space projection (AR-NSP) scheme is proposed for secure transmission with artificial noise (AN)-aided directional modulation (DM) in wireless networks. The proposed scheme is carried out in three steps. Firstly, the directions of arrival (DOAs) of the signals from  the  desired user and eavesdropper
are estimated by the Root Multiple Signal Classificaiton (Root-MUSIC) algorithm and the related signal-to-noise ratios (SNRs) are estimated based on the ratio of the corresponding eigenvalue to the minimum eigenvalue of the covariance matrix of the received signals. In the second step, the value intervals of DOA estimation errors are predicted based on the DOA and SNR estimations. Finally, a robust NSP beamforming DM system is designed according to the afore-obtained estimations and predictions. Our examination shows that the proposed scheme can significantly outperform the conventional non-adaptive robust scheme and non-robust NSP scheme in terms of achieving a much lower bit
error rate (BER) at the desired user and a much higher secrecy rate (SR). In addition, the BER and SR performance gains achieved by the proposed scheme relative to other schemes increase with the value range of DOA estimation error.
\end{abstract}




\IEEEpeerreviewmaketitle

%
%
%
%
\section{Introduction}
Due to the broadcast nature of wireless mediums, private information is vulnerable to be intercepted by unintended users, which causes ever-increasing concerns on the security of wireless networks. As such, transmission, storage, processing, and protection of confidential information in wireless networks are of growing research interests \cite{Elisa,Cong,Huili,Fenghua}. Against this background,, there is an increasing demand for designing secure transmissions to protect the legitimate users against being overheard \cite{Wyner1975,Zou2016,Fei,Trappe,Chen,Goel,zhao,YAN,Yang,Hu2017,Zou2015_1,Hui,Lun,kalantari}. In \cite{Wyner1975}, Wyner conducted pioneering research on a discrete memoryless wiretap channel for secure communication in the presence of an eavesdropper and proposed the notion of secrecy capacity. Different from the conventional cryptographic technologies, of which the achieved security is based on the high computational complexity, physical layer security (PLS) can be utilized to provide an ever-lasting information-theoretic security. In addition, PLS does not require cryptographic keys in order to achieve secrecy, which eliminate the complicated key distributions and managements. Consequently, it is essential to take advantage of PLS to defend unintended user against eavesdropping confidential information in the upcoming fifth generation (5G) and beyond wireless networks \cite{Trappe}.

A secure transmission scheme was proposed, using the basic idea of original symbol phase rotated (OSPR) in \cite{Chen}, to randomly rotate the phase of original symbols at base station (BS) before they are transmitted, so that the legitimate receivers were able to successfully recover original symbols by taking proper inverse operation while the illegitimate users could not detect most of original symbols. Artificial noise (AN)-aided security transmission was employed in \cite{Goel,zhao,YAN,Yang,Hu2017} to disrupt the unauthorized users from tapping the confidential messages. In \cite{Goel}, provided that the channel state information (CSI) was publicly and perfectly known, the transmitter produced AN to guarantee secure communication in two scenarios, the transmitter equipped with multiple antennas and multiple-cooperative-relays-aided  secure transmission. Cooperative relays and symbol-level precoding were respectively considered in \cite{Zou2015_1,Hui,Lun} and \cite{kalantari} to enhance PLS and prevent eavesdropping attacks in wireless networks. Moreover, security and privacy is critical in the field of internet of things (IoT), and PLS plays a key role in IoT networks \cite{Stankovic,Mukherjee,Qian,LinHu,Ren}. The author in \cite{Mukherjee} summarized the low-complexity PLS schemes for IoT and presented some new research topics in the context of future IoT. In \cite{Qian}, randomize-and-forward relay strategy was leveraged for secure relay communication without the knowledge of number and locations of eavesdroppers in IoT. And all devices equipped with single antenna and multiple antennas were analyzed, respectively. To enhance the PLS, the authors in \cite{LinHu} investigated secure downlink transmission with the help of cooperative jamming in the IoT networks. Additionally, wireless physical layer identification (WPLI) system was proposed in \cite{Ren} to increase the user capacity by assigning multiple devices to one user in IoT networks.

Directional modulation (DM),  as a critical secure wireless technique,  attracts tremendous  research interests from both academia and industry and plays an increasingly important role in the context of PLS of wireless communications and networks. An orthogonal vector approach, using the concept of null-space projection (NSP), was proposed in \cite{Ding2014} to allow analysis and synthesis of DM transmitter arrays. In \cite{Tao}, an AN-aided zero-forcing method was proposed, which was intimately related to the concept of pseudo-inverse of matrix, where the dynamic characteristic of DM was achieved by randomly changing the AN vector at the symbol-level rate. However, it was assumed that the perfect knowledge of the desired direction was available. Additionally, maximizing the so-called signal-to-leakage-and-noise ratio (SLNR) for all users simultaneously to design precoders was proposed  in \cite{Sadek2007,Tong}. In \cite{Hu2016,Shu2016,Zhu}, the direction angle estimation error was taken into account and a robust method of designing the beamforming vector of confidential information and projection matrix of AN was presented. The authors of \cite{Hu2016} derived a closed-form expression for the null space of conjugate transpose of the steering vector in the desired direction and proposed a robust DM synthesis method based on conditional minimum mean square error (MMSE) criterion. In \cite{Shu2016}, the authors extended  the idea in  \cite{Hu2016} to the broadcasting scenario, where the conditional leakage beamforming scheme was presented.  In \cite{Zhu},  without the knowledge of direction measurement error,  a blind robust secure leakage method of using main-lobe-integration was constructed for multi-user multiple input multiple output (MIMO) scenario. Furthermore, the authors extended the application of DM to a multi-cast scenario in \cite{Xu}. In this work, two methods,  maximum group receive power plus null-space projection  scheme and Max-SLNR  plus maximum-AN-leakage-and-noise ratio scheme, were proposed  to enhance PLS and improve the secrecy sum-rate compared with block diagonalization using the same order low computational complexity.

All the aforementioned research works on DM only focus on how to design  beamforming schemes with perfect information of the direction angle or imperfect such information but with given error distributions. We note that it is hard, if not impossible, to adapt the DM techniques proposed in the aforementioned works as per different channel conditions, such as the varying signal-to-noise ratio (SNR). Actually, in all the robust DM beamforming schemes, the first step is to estimate the direction-of-arrival (DOA) together with the SNR and predict the value range of DOA estimation error at DM transmitter, which is mandatory in a practical DM network. However, this step is overlooked in the context of designing robust synthesis methods of DM in the literature. In addition, due to the randomness of wireless channels, the transmitter may have to estimate the DOA and SNR periodically. To address the estimation problem in DM networks, in this work we propose an adaptive and robust NSP (AR-NSP) method, which consists of three steps. Firstly, the DOAs are measured by Root Multiple Signal Classificaiton (Root-MUSIC) rule and the corresponding SNRs are estimated. Then the value ranges of the DOA estimation error are predicted. In the last step, a robust NSP scheme is proposed to design the beamforming vector of confidential information and the AN projection vector without the statistical knowledge of channel noise. Our main contributions are summarized as below.
\begin{enumerate}
  \item  Based on the singular value decomposition (SVD) of the covariance matrix of the receive signals achieved by the Root-MUSIC method, we first estimate the receive SNRs for the desired and eavesdropping users. These estimated SNRs are further utilized to predict the value intervals of DOA estimation errors with the aid of  the Cramer-Rao lower bound (CRLB) expression of unbiased DOA estimators.

 \item  As per the predicted value intervals of DOA estimation errors, we propose an AR-NSP  scheme for DM, aiming to enhance the secure transmission of confidential information. The proposed method can adapt its beamforming vector of the confidential information and the projection vector of AN according to the quality of wireless channels. Our examine shows that the proposed method can achieve noticeable performance gains relative to the non-adaptive robust conditional MMSE method presented in \cite{Hu2016} and the non-robust NSP method proposed in \cite{Ding2014} in terms of achieving a much lower bit error rate (BER) and a significantly higher security rate (SR).
\end{enumerate}

The remainder of this paper is organized as follows. Section II presents the DM system model. In Section III, DOA and SNR are first estimated and then the value intervals of DOA estimation errors are predicted. In addition, our proposed AR-NSP  scheme is detailed in this section. Simulation and numerical results are presented in Section IV. Finally, we draw our conclusions in Section V.

\emph{Notations:} Throughout the paper, matrices, vectors, and scalars are denoted by letters of bold upper case, bold lower case, and lower case, respectively. Signs $(\cdot)^T$, $(\cdot)^H$, and $\parallel\cdot\parallel$ represent transpose, conjugate transpose, and norm, respectively. The notation $\mathbb{E}\{\cdot\}$ denotes the expectation operation. $\textbf{I}_N$ denotes the $N\times N$ identity matrix.

\section{System Model}
\begin{figure}[tp]
\centering
\subfigure[System model of the first time slot ]{
\label{Fig1_Sys_Mod_a}
\includegraphics[width=0.6\columnwidth]{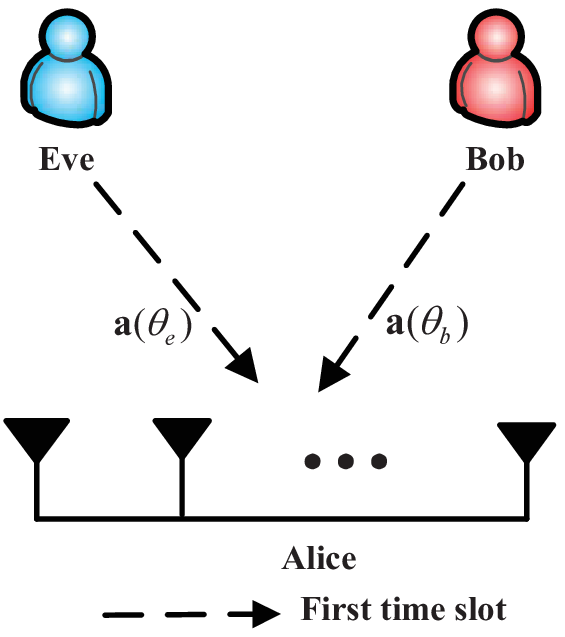}}
\hspace{1in}
\subfigure[System model of the second time slot]{
\label{Fig1_Sys_Mod_b}
\includegraphics[width=0.6\columnwidth]{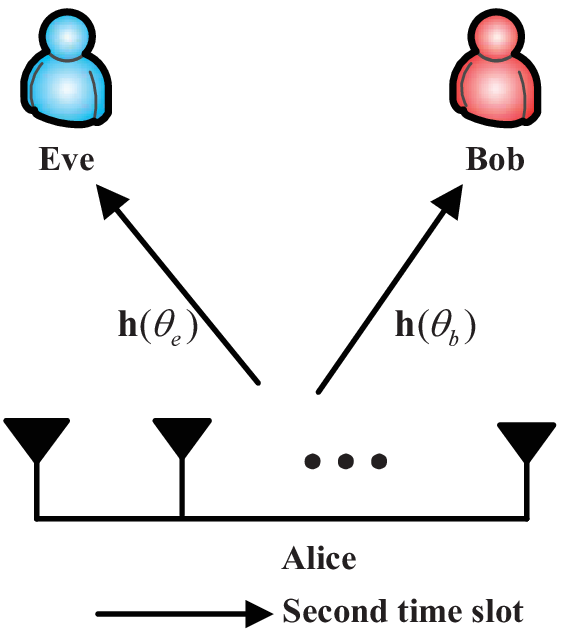}}
\caption{System model of adaptive and robust directional modulation.}
\label{Sys_Mod}
\end{figure}

The system model of interest is shown in Fig.~\ref{Sys_Mod}, where Alice is a transceiver equipped with an $N$-antenna array, while each of Bob and Eve is equipped with a single antenna. The secure transmission in the considered system model incurs in two consecutive time slots. In the first time slot, Bob and Eve transmit signals, while Alice receives their signals. Alice measures Bob's and Eve's directions and predicts their value intervals of the direction measurement errors. In the second time slot, Alice uses the predicted directions and error value intervals to design the AR-NSP scheme. Subsequently, Alice transmits the confidential information to Bob using the AR-NSP scheme in order to prevent Eve from eavesdropping the confidential information successfully.

\subsection{System Model in the First Time Slot}
In the first time slot, Bob and Eve are the transmitters,  while Alice is the receiver. In fact, Bob transmits useful signals to Alice, while Eve is an active attacker who would like to intercept the communications between Alice and Bob. By receiving the transmitted signals from Bob and  Eve,  Alice can measure their directions. Bob and Eve send the passband signals $s_{b}(t)e^{j2\pi f_{b} t}$ and $s_{e}(t)e^{j2\pi f_{e} t}$, respectively, where $s_{b}(t)$, and $s_{e}(t)$ represent their corresponding baseband signals, and $f_b$ and $f_e$ are carrier frequencies, respectively. With the narrow-band assumption, the receive signal at Alice can be expressed as
\begin{equation}
\mathbf{r}(t)=s_b(t)\mathbf{a}(\theta_b)+s_e(t)\mathbf{a}(\theta_e)+\mathbf{n}(t),
\end{equation}
where $\mathbf{a}(\theta_b)$ and $\mathbf{a}(\theta_e)$ are the array manifolds for Bob and Eve defined by
\begin{align}\label{vector_b}
\mathbf{a}(\theta_b)=\left[1, e^{j\frac{2\pi}{\lambda_b}d\sin(\theta_b)}, \cdots, e^{j\frac{2\pi}{\lambda_b}(N-1)d\sin(\theta_b)}\right]^T,
\end{align}
and
\begin{align}\label{vector_e}
\mathbf{a}(\theta_e)=\left[1, e^{j\frac{2\pi}{\lambda_e}d\sin(\theta_e)}, \cdots, e^{j\frac{2\pi}{\lambda_e}(N-1)d\sin(\theta_e)}\right]^T,
\end{align}
respectively, and $d$ denotes the spacing between two adjacent elements, while $\lambda_b=c/f_b$ and $\lambda_e=c/f_e$ are the wavelengths of the transmit signal carriers from Bob and Eve, respectively.

After down-conversion and A/D operations, the resultant sampled receive baseband  signal can be represented as
\begin{equation}\label{Uplink-Rx-Signal}
\mathbf{r}[k]=s_b[k]\mathbf{a}(\theta_b)+s_e[k]\mathbf{a}(\theta_e)+\mathbf{n}[k], k=1,2,\cdots N_s,
\end{equation}
where $N_s$ is the number of snapshots, $P_b=\mathbb{E}\{\mid s_b[k]\mid^2\}$ and $P_e=\mathbb{E}\{\mid s_e[k]\mid^2\}$ are the transmit powers of Bob and Eve, and $\mathbf{n}[k]$ is the receiving noise vector with distribution $\mathcal{C}\mathcal{N}(\mathbf{0},\hat\sigma^2\mathbf{I}_N)$.

\subsection{System Model in the Second Time Slot}
Given that Alice has estimated the direction angles of Bob and Eve in the first time slot,  she transmits its confidential messages by the precoding vector $\mathbf{v}_b$ towards Bob, and prevents Eve from intercepting Alice's confidential messages by using AN projection vector $\mathbf{w}_e$ in the second time slot. The transmit signal of Alice is given by
\begin{equation}\label{Tx signal s}
\mathbf{s}_m=\beta_1\sqrt{P_s}\mathbf{v}_bx_m+\beta_2\sqrt{P_s}\mathbf{w}_ez_m,
\end{equation}
where $P_s$ is the total transmission power of Alice, $\beta_1$ and $\beta_2$ are the power allocation (PA) coefficients of confidential messages and AN with $\beta^2_1+\beta^2_2=1$, respectively, $\mathbf{v}_b\in\mathbb{C}^{N\times1}$denotes the transmitting beamforming vector of useful signals for forcing the confidential messages to the desired direction and $\mathbf{w}_e\in\mathbb{C}^{N\times 1}$ is the AN projection vector of leading AN to the undesired direction, $x_m$ is the confidential message with $\mathbb{E}\left\{x^H_mx_m\right\}=1$, and $z_m$ denotes the scalar AN varying from one symbol to another with $\mathbb{E}\left\{z^H_mz_m\right\}=1$.

We suppose that $\mathbf{s}_m$ passes through the line-of-sight (LOS) channel and then the receive signal at direction $\theta$ is given by
\begin{align}\label{Rx_signal y}
y(\theta)
&=\mathbf{h}^{H}(\theta)\mathbf{s}_m+n_r\nonumber\\
&=\beta_1\sqrt{P_s}\mathbf{h}^{H}(\theta)\mathbf{v}_bx_m+\beta_2\sqrt{P_s}\mathbf{h}^{H}(\theta)\mathbf{w}_ez_m+n_r,
\end{align}
where $\theta$ is the direction angle between transmitter and receiver and $\theta\in[-\pi/2,\pi/2]$, $n_r$ is additive white Gaussian noise (AWGN) of obeying the complex Gaussian distribution $\mathcal{C}\mathcal{N}(0,\sigma_r^2)$, $\mathbf{h}(\theta)\in\mathbb{C}^{N\times1}$ is the array manifold given by
\begin{equation}\label{h_theta}
\mathbf{h}(\theta)=\frac{1}{\sqrt{N}}\left[e^{j2\pi\Psi_{\theta}(1)}, \cdots, e^{j2\pi\Psi_{\theta}(n)}, \cdots, e^{j2\pi\Psi_{\theta}(N)}\right]^T
\end{equation}
and the phase function $\Psi_{\theta}(n)$ along direction $\theta$ can be written as
\begin{equation}\label{var_phi}
\Psi_{\theta}(n)\triangleq-\frac{(n-(N+1)/2)d\cos\theta}{\lambda}, n=1,2,\cdots N.
\end{equation}
As such, the receive signals at Bob and Eve can be expressed as
\begin{align}\label{Rx_signal bob}
y(\theta_b)
&=\mathbf{h}^{H}(\theta_b)\mathbf{s}_m+n_b\nonumber\\
&=\beta_1\sqrt{P_s}\mathbf{h}^{H}(\theta_b)\mathbf{v}_bx_m+\beta_2\sqrt{P_s}\mathbf{h}^{H}(\theta_b)\mathbf{w}_ez_m+n_b,
\end{align}
and
\begin{align}\label{Rx_signal eve}
y(\theta_e)
&=\mathbf{h}^{H}(\theta_e)\mathbf{s}_m+n_e\nonumber\\
&=\beta_1\sqrt{P_s}\mathbf{h}^{H}(\theta_e)\mathbf{v}_bx_m+\beta_2\sqrt{P_s}\mathbf{h}^{H}(\theta_e)\mathbf{w}_ez_m+n_e,
\end{align}
respectively, where $n_b\sim\mathcal{C}\mathcal{N}(0,\sigma_b^2)$ and $n_e\sim\mathcal{C}\mathcal{N}(0,\sigma_e^2)$ represent the AWGN at Bob and Eve, respectively. Without loss of generality, we assume that $\sigma_r^2=\sigma_b^2=\sigma_e^2=\sigma^2$.
Following (\ref{Rx_signal bob}) and (\ref{Rx_signal eve}), we have the achievable  rates at Bob and Eve as
\begin{align}\label{Rb}
R(\theta_b)
 & \triangleq I(y(\theta_b);[x,\theta_b])\nonumber\\
 & = \log_2\left(1+\frac{\beta_1^2P_s \mathbf{h}^H(\theta_b)\mathbf{v}_b\mathbf{v}_b^H\mathbf{h}(\theta_b)}{\sigma^2_b+\beta_2^2P_s \mathbf{h}^H(\theta_b)\mathbf{w}_e\mathbf{w}_e^H\mathbf{h}(\theta_b)}\right)
\end{align}
and
\begin{align}\label{Re}
R(\theta_e)
 & \triangleq I(y(\theta_e);[x,\theta_e])\nonumber\\
 & = \log_2\left(1+\frac{\beta_1^2P_s \mathbf{h}^H(\theta_e)\mathbf{v}_b\mathbf{v}_b^H\mathbf{h}(\theta_e)}{\sigma^2_e+\beta_2^2P_s \mathbf{h}^H(\theta_e)\mathbf{w}_e\mathbf{w}_e^H\mathbf{h}(\theta_e)}\right),
\end{align}
which yield  the achievable secrecy rate $R_s$ as
\begin{equation}\label{Rs}
R_s=\max\left\{0,R(\theta_b)-R(\theta_e)\right\}.
\end{equation}

\section{Proposed Adaptive and Robust NSP Scheme}
In this section, we present the proposed AR-NSP scheme to design the beamforming vector $\mathbf{v}_b$ and projection vector $\mathbf{w}_e$. Firstly,  the direction angles, including the directions of desired user and eavesdropper, and the corresponding SNRs are  estimated using the Root-MUSIC algorithm. Subsequently, based on the estimated angles and SNRs, the value intervals of the DOA estimation errors are predicted with the aid of CRLB.  Finally, the AR-NSP scheme is constructed to enhance the secrecy performance of the considered DM system.

\subsection{DOA and SNR Estimations with Error Range Prediction}

In the first time slot, by collecting  the receive vector $\mathbf{r}[k]$ of $N_s$ snapshots for $\forall k\in\left\{1,~2,~\cdots, N_s\right\}$,   the covariance matrix of $\mathbf{r}[k]$ in $(\ref{Uplink-Rx-Signal})$ is given by
\begin{align}\label{Ryy}
\mathbf{R}_{yy}
&=\frac{1}{N_s}\sum_{k=1}^{N_s}\mathbf{r}[k]\mathbf{r}^H[k]\nonumber\\
&=\frac{1}{N_s}\sum_{k=1}^{N_s}\{P_b\mathbf{a}(\theta_b)\mathbf{a}^H(\theta_b)+P_e\mathbf{a}(\theta_e)\mathbf{a}^H(\theta_e)\}\nonumber\\
&+\frac{1}{N_s}\sum_{k=1}^{N_s}\hat\sigma^2\mathbf{I}_N.
\end{align}
As per the SVD operation, $\mathbf{R}_{yy}$ can be reformulated as follows
\begin{align}\label{covariance matrix}
\mathbf{R}_{yy}
&=\mathbf{U}\mathbf{\Lambda}\mathbf{U}^H+\hat\sigma^2\mathbf{I}_N\nonumber\\
&=\mathbf{U}\mathbf{\Pi}\mathbf{U}^H,
\end{align}
where $\mathbf{U}$ is an unitary matrix, and $\mathbf{\Pi}$ is a diagonal matrix given by
\begin{equation}\label{SVD}
\mathbf{\Pi}=\mathbf{diag}(NP_b+\hat\sigma^2, NP_e+\hat\sigma^2, \hat\sigma^2, \cdots, \hat\sigma^2).
\end{equation}

Referring to the Root-MUSIC in \cite{Gross2005}, the estimated value of  direction angle $\hat\theta_m$ can be computed as per the following expression
\begin{equation}\label{theta_m_est}
\hat\theta_m=\arcsin\left(\frac{\arg(z_m)}{ld}\right),  m\in \{1, 2\},
\end{equation}
where
\begin{equation}\label{k}
l=\frac{2\pi}{\lambda},
\end{equation}
and $m=1$ means $\theta_1=\theta_b$, i.e., the desired direction,  $m=2$  means that $\theta_2=\theta_e$, the eavesdropper direction,  and $z_m$ are the two roots of the polynomial of the Root-MUSIC in \cite{Gross2005}, which are the two solutions closet to the unit circle inside the unit circle. By performing the SVD operation on the covariance matrix $\mathbf{R}_{yy}$, we can obtain the signal subspace $\mathbf{E}_S$, the noise subspace $\mathbf{E}_N$,  and the associated eigenvalues in descending order, i.e,
\begin{equation}\label{EVD}
\mu_1>\mu_2>\cdots>\mu_{M+1}>\cdots>\mu_N,
\end{equation}
where $M=2$ is the number of incoming signals. Then the noise variance is given by
\begin{equation}\label{sigma_m}
\hat\sigma^2=\frac{\mu_{M+1}+\cdots+\mu_N}{N-M},
\end{equation}
which yields the SNR estimated value for emission source $m$ as
\begin{equation}\label{SNR}
\hat\gamma_m=\frac{\mu_m-\hat\sigma^2}{N\hat\sigma^2},
\end{equation}
which can be used to predict the intervals of the DOA estimation errors. From the definition of CRLB, we have the variance of direction angle error  is lower bounded by \cite{Tuncer2009}
\begin{equation}\label{CRLB}
\sigma^2_{\theta_m}\ge\mathrm{CRLB}_m\approx\frac{\lambda^2}{8\pi^2N_s\hat\gamma_m\cos^2\hat\theta_m L},
\end{equation}
where
\begin{equation}\label{d}
L=\sum_{n=1}^Nd^2_n,
\end{equation}
and
\begin{equation}\label{dn}
d_n=(n-\frac{N+1}{2})d, n=1,2,\cdots N.
\end{equation}
Since we can get the lower bound of standard deviation $\sigma_{\theta_m}$, the approximated value interval of DOA estimation error is given by
\begin{equation}\label{ang_interval}
\theta_m\in \left[\hat\theta_m-{\Delta\hat\theta}_{m, max},~\hat\theta_m+{\Delta\hat\theta}_{m, max}\right]
\end{equation}
where
\begin{equation}\label{angel error}
{\Delta\hat\theta}_{m, max}=\rho~\sigma_{\theta_m}\approx\rho\sqrt{(\mathrm{CRLB}_m)}.
\end{equation}
By increasing the value of $\rho$ in (\ref{angel error}), the probability that (\ref{ang_interval}) holds tends to one. We note that here we adopt $\rho = 1$ and in the numerical section we examine the impact of $\rho$ on the secrecy performance of the proposed AR-NSP scheme.

\subsection{Design of the AR-NSP Scheme}
In (\ref{theta_m_est}), the estimated direction angle can be expanded as
\begin{align}
\hat{\theta}_m=\theta_m+\Delta\theta_m,
\end{align}
where $\theta_m$ is the ideal direction angle, and $\Delta\theta_m$ stands for the angle error. Combining (\ref{ang_interval}) and the above expression, we have
\begin{equation}\label{ang_err_interval}
\Delta\theta_m \in \left[-{\Delta\hat\theta}_{m, max},~{\Delta\hat\theta}_{m, max}\right].
\end{equation}
 Additionally, for convenience of deriving below, the probability density function (PDF) of angle error is approximated to be a uniform distribution given by
\begin{equation}\label{pdf}
   p\left(\Delta\theta_m\right)=\left\{
   \begin{aligned}
    &\frac{1}{2\Delta\theta_m}, &-{\Delta\hat\theta}_{m, max}\leq\Delta\theta_m\leq{\Delta\hat\theta}_{m, max},\\
    &0,                       &\mathrm{otherwise},
   \end{aligned}
   \right.
\end{equation}
where ${\Delta\hat\theta}_{m, max}$ denotes the interval of DOA estimation error.

Given the estimated direction angles $\hat\theta_b$ and $\hat\theta_e$, in accordance with \cite{Lun} and \cite{Tomas2000},  we have  the conditional expectation of robust beamforming vector of the confidential information and the AN projection vector as
\begin{align}\label{v1}
\mathbb{E}[\bar{\mathbf{v}}_b\mid\hat{\theta}_b,\hat{\theta}_e]
&=\mathbb{E}[(\mathbf{I}_{N}-\mathbf{h}({\theta}_e)\mathbf{h}^H({\theta}_e))\mathbf{h}({\theta}_b)\mid\hat{\theta}_b,\hat{\theta}_e]\nonumber\\
&=\left(\mathbf{I}_N-\mathbf{R}_e\right)\mathbf{u}_b,
\end{align}
and
\begin{align}\label{w1}
\mathbb{E}[\bar{\mathbf{w}}_e\mid\hat{\theta}_b,\hat{\theta}_e]
&=\mathbb{E}[(\mathbf{I}_{N}-\mathbf{h}({\theta}_b)\mathbf{h}^H({\theta}_b))\mathbf{h}({\theta}_e)\mid\hat{\theta}_b,\hat{\theta}_e]\nonumber\\
&=\left(\mathbf{I}_N-\mathbf{R}_b\right)\mathbf{u}_e,
\end{align}
respectively, where we define
\begin{align}
&\mathbf{R}_b\triangleq\mathbb{E}[\mathbf{h}({\theta}_b)\mathbf{h}^H({\theta}_b)\mid\hat\theta_b]\nonumber\\
&\mathbf{R}_e\triangleq\mathbb{E}[\mathbf{h}({\theta}_e)\mathbf{h}^H({\theta}_e)\mid\hat\theta_e]\nonumber\\
&\mathbf{u}_b\triangleq\mathbb{E}[\mathbf{h}({\theta}_b)\mid\hat\theta_b]\nonumber\\
&\mathbf{u}_e\triangleq\mathbb{E}[\mathbf{h}({\theta}_e)\mid\hat\theta_e],
\end{align}
and the corresponding normalized beamforming vector $\mathbf{v}_b$ and projection vector $\mathbf{w}_e$ can be written as
\begin{equation}\label{beamforming vector}
\mathbf{v}_b=\frac{\mathbb{E}[\bar{\mathbf{v}}_b\mid\hat{\theta}_b,\hat{\theta}_e]}{\parallel\mathbb{E}[\bar{\mathbf{v}}_b\mid\hat{\theta}_b,\hat{\theta}_e]\parallel},
\end{equation}
and
\begin{equation}\label{projection vector}
\mathbf{w}_e=\frac{\mathbb{E}[\bar{\mathbf{w}}_e\mid\hat{\theta}_b,\hat{\theta}_e]}{\parallel\mathbb{E}[\bar{\mathbf{w}}_e\mid\hat{\theta}_b,\hat{\theta}_e]\parallel}.
\end{equation}
Since the estimated direction angles and their intervals of DOA estimation errors are known, the derivation of $p$-th row and $q$-th column entry of matrix $\mathbf{R}_e$ is given by
\begin{align}\label{R1}
{R}_e(p,q)
&=\mathbb{E}[\mathbf{h}_{p}({\hat\theta}_e-\Delta\theta_e)\mathbf{h}^H_{q}({\hat\theta}_e-\Delta\theta_e)]\nonumber\\
&=\frac{1}{N}\int_{-{\Delta\hat\theta}_{e,max}}^{\Delta\hat\theta_{e,max}}e^{j\pi(q-p)\cos(\hat\theta_e-\Delta\theta_e)}\nonumber\\
&\times p\left(\Delta\theta_e\right)d(\Delta\theta_e).
\end{align}
In order to simplify the presentation of (\ref{R1}), we define
\begin{align}
a_{pq}&=j\pi(q-p)\cos(\hat\theta_e),\label{a_pq}\\
b_{pq}&=j\pi(q-p)\sin(\hat\theta_e),\label{b_pq}
\end{align}
and define a new variable as
\begin{equation}\label{x_e}
x_e=\frac{\Delta\theta_e}{c_e},
\end{equation}
where
\begin{equation}\label{c_e}
c_e=\Delta\hat\theta_e,_{max}\pi^{-1}.
\end{equation}
Using the above definition, (\ref{R1}) can be reformulated as
\begin{align}\label{R2}
{R}_e(p,q)
&=\frac{1}{2\pi N}\int_{-\pi}^{\pi}e^{a_{pq}\cos(c_ex_e)+b_{pq}\sin(c_ex_e)}dx_e\nonumber\\
&=\frac{1}{N}\textrm{g}_I(a_{pq},b_{pq},c_e),
\end{align}
where
\begin{equation}\label{g}
\textrm{g}_I(a_{pq},b_{pq},c_e)=\frac{1}{2\pi}\int_{-\pi}^{\pi}e^{a_{pq}\cos(c_ex_e)+b_{pq}\sin(c_ex_e)}dx_e.
\end{equation}
Likewise, we can derive the $n$-th element of vector $\mathbf{u}_b$ as
\begin{align}\label{u1}
{u}_b(n)
&=\mathbb{E}[\mathbf{h}_{n}({\hat\theta}_b-\Delta\theta_b)]\nonumber\\
&=\frac{1}{\sqrt{N}}\int_{-\Delta\hat\theta_b,_{max}}^{\Delta\hat\theta_b,_{max}}e^{-j\pi(n-\frac{N+1}{2})\cos({\hat\theta}_b-\Delta\theta_b)}\nonumber\\
&\times p(\Delta\theta_b)d(\Delta\theta_b)\nonumber\\
&=\frac{1}{2\pi\sqrt{N}}\int_{-\pi}^{\pi}e^{e_n\cos(c_bx_b)+f_n\sin(c_bx_b)}dx_b\nonumber\\
&=\frac{1}{\sqrt{N}}\textrm{g}_I(e_n,f_n,c_b),
\end{align}
where
\begin{equation}\label{e_n}
e_n=-j\pi(n-\frac{N+1}{2})\cos(\hat\theta_b),
\end{equation}
\begin{equation}\label{f_n}
f_n=-j\pi(n-\frac{N+1}{2})\sin(\hat\theta_b),
\end{equation}
\begin{equation}\label{x_b}
x_b=\frac{\Delta\theta_b}{c_b},
\end{equation}
\begin{equation}\label{c_b}
c_b=\Delta\hat\theta_b,_{max}\pi^{-1}.
\end{equation}
Thus, the corresponding entry of matrix $\mathbf{R}_b$ and vector $\mathbf{u}_e$ can be computed in the same manner.

\begin{figure}[tp]
 \centering
 \includegraphics[width=0.36\textwidth]{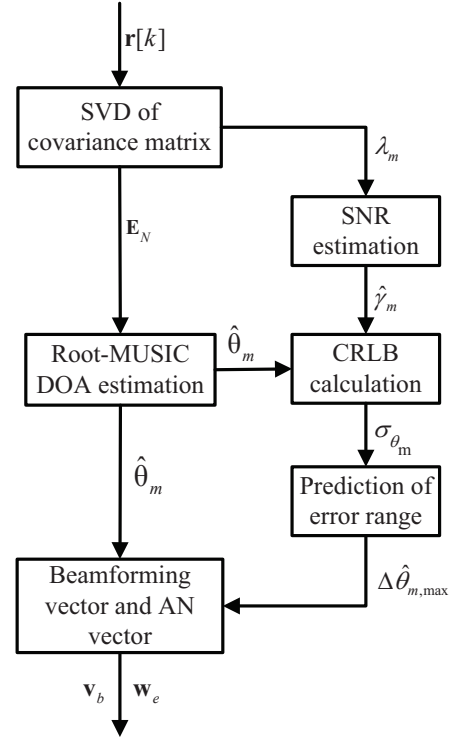}\\
 \caption{Block diagram of the proposed AR-NSP scheme.}\label{Struc}
\end{figure}

Taking the above two subsections into account, Fig.~\ref{Struc} presents the flow chart for the proposed AR-NSP scheme. At the same time,  we also summarize the detailed operational procedures of our proposed AR-NSP scheme in Algorithm~1.

\begin{algorithm}[h]
\begin{enumerate}
  \item Firstly,  Bob and Eve act as transmitters and Alice (as the receiver) measures the Bob/desired and Eve/eavesdropper direction angles via Root-MUSIC.
  \item   The receive SNR values for Bob and Eve  at Alice are estimated using the singular value ratio of SVD in Root-MUSIC.
   \item   Using $\mathrm{CRLB}_m$ on the basis of estimated direction angle $\hat\theta_m$ and estimated SNR $\hat\gamma_m$, the intervals of DOA estimation errors are approximately predicted.
   \item   The probability density functions of angle errors  are approximated as uniform distributions over the predicted value ranges.
   \item  Finally,  Alice (as the transmitter) adopts the proposed AR-NSP beamforming method, to design the beamforming vector $\mathbf{v}_b$ of confidential messages and projection vector $\mathbf{w}_e$ of AN based on the results from Step 1) to Step 4), and transmits confidential information to Bob.
\end{enumerate}
\caption{Proposed AR-NSP algorithm}\label{algorithm 1}
\end{algorithm}

\section{Numerical Results}
In this section, we numerically examine the secrecy performance of the proposed adaptive and robust NSP scheme.  Without other statements, the adopted system parameters are set as follows: quadrature phase shift keying(QPSK) modulation is adopted, the spacing between two adjacent antennas is set as $d=\lambda/2$, the total transmitting power is set as $P_s=30$dBm, PA factors are set as $\beta_1^2=0.9$ and $\beta_2^2=0.1$, the actual desired direction is set as $\theta_b=45^{\circ}$, and the actual eavesdropping direction is set as $\theta_e=-30^{\circ}$.

\begin{figure}[h]
  \centering
  \includegraphics[width=0.52\textwidth]{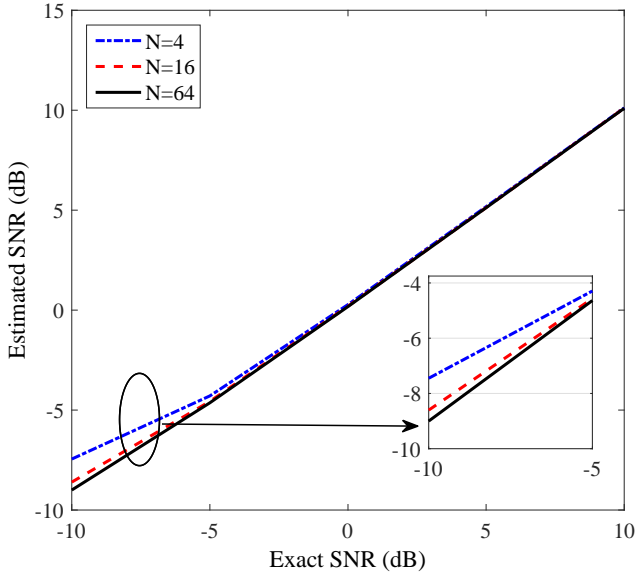}\\
  \caption{Estimated SNR versus exact SNR with different numbers of antennas.}\label{SNR_EST}
\end{figure}

\begin{figure}[h]
  \centering
  \includegraphics[width=0.52\textwidth]{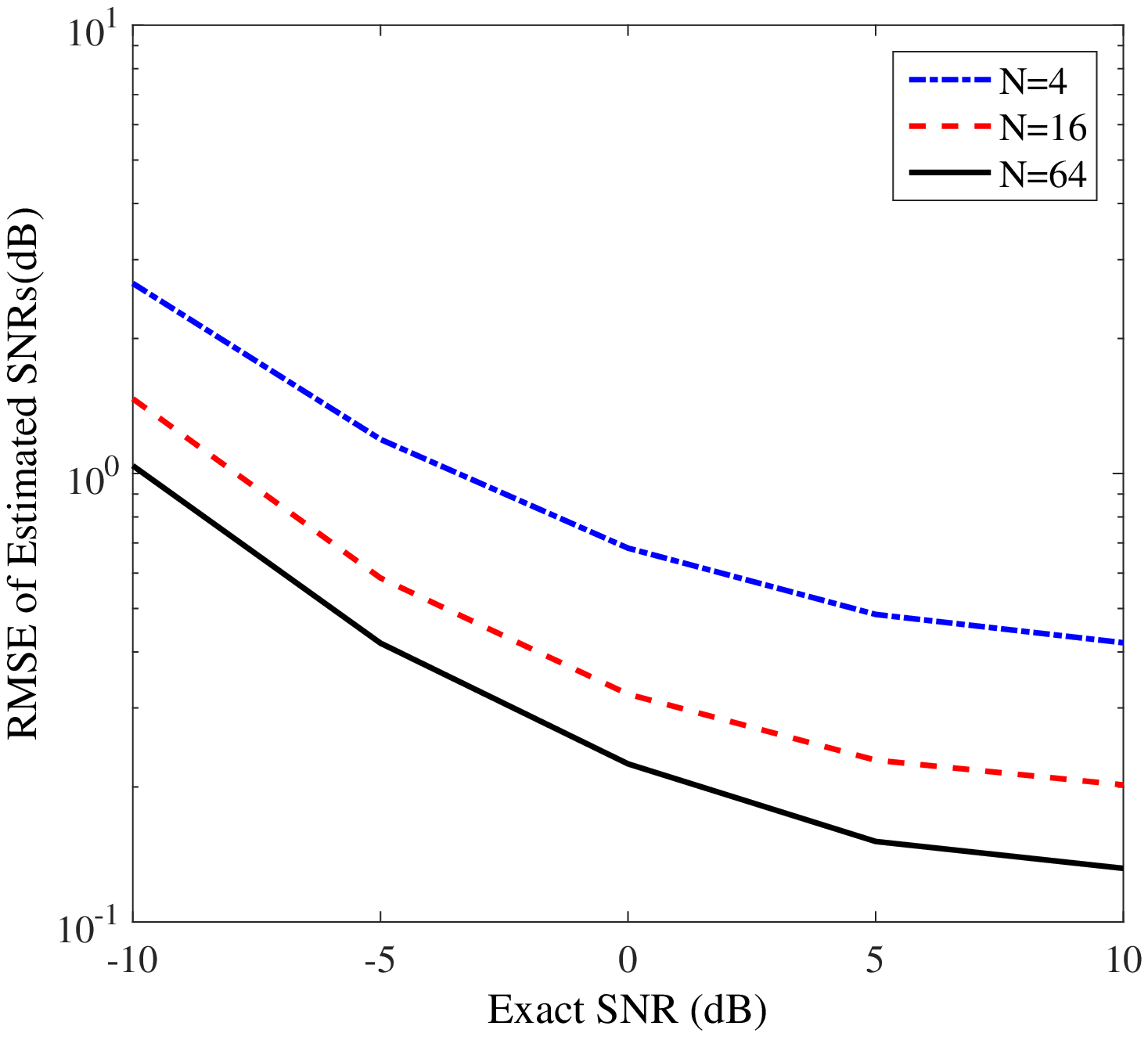}\\
  \caption{Root mean squared error of the SNR estimation versus exact SNR with different numbers of antennas.}\label{RMSE_SNR}
\end{figure}

\begin{figure}[h]
  \centering
  \includegraphics[width=0.52\textwidth]{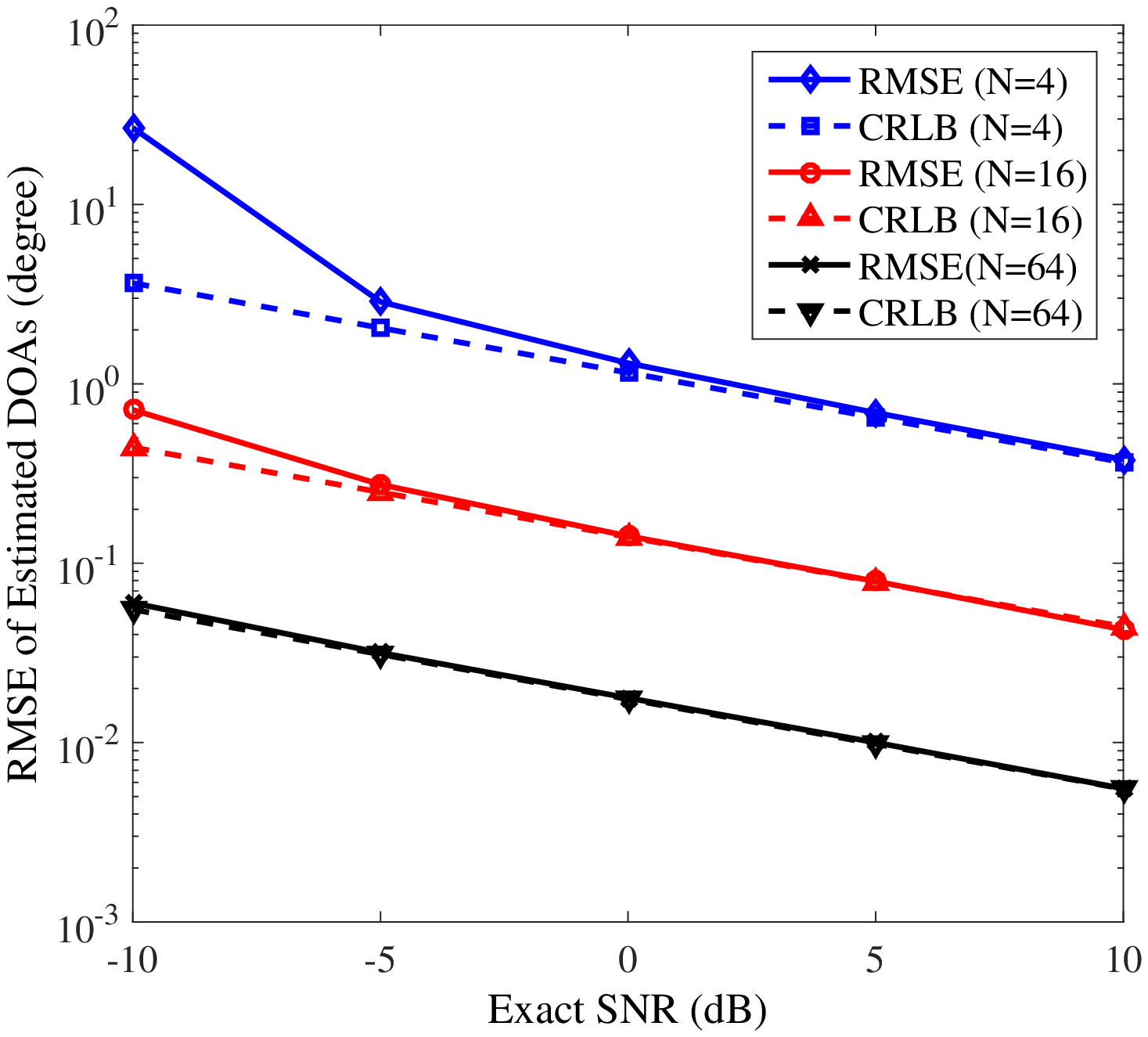}\\
  \caption{Root mean squared error of the estimation of $\theta$ versus exact SNR with different numbers of antennas.}\label{CRB_SNR}
\end{figure}

\begin{figure}[h]
  \centering
  \includegraphics[width=0.52\textwidth]{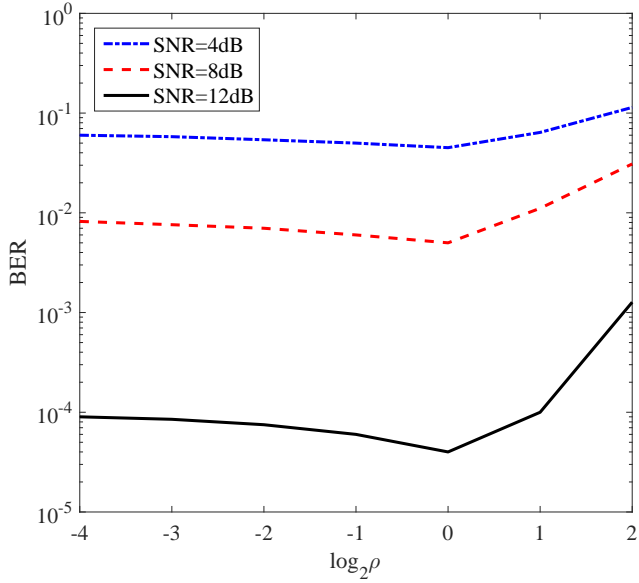}\\
  \caption{BER versus different values of $\rho$ with different SNRs.}\label{rho_BER}
\end{figure}

\begin{figure}[h]
  \centering
  \includegraphics[width=0.52\textwidth]{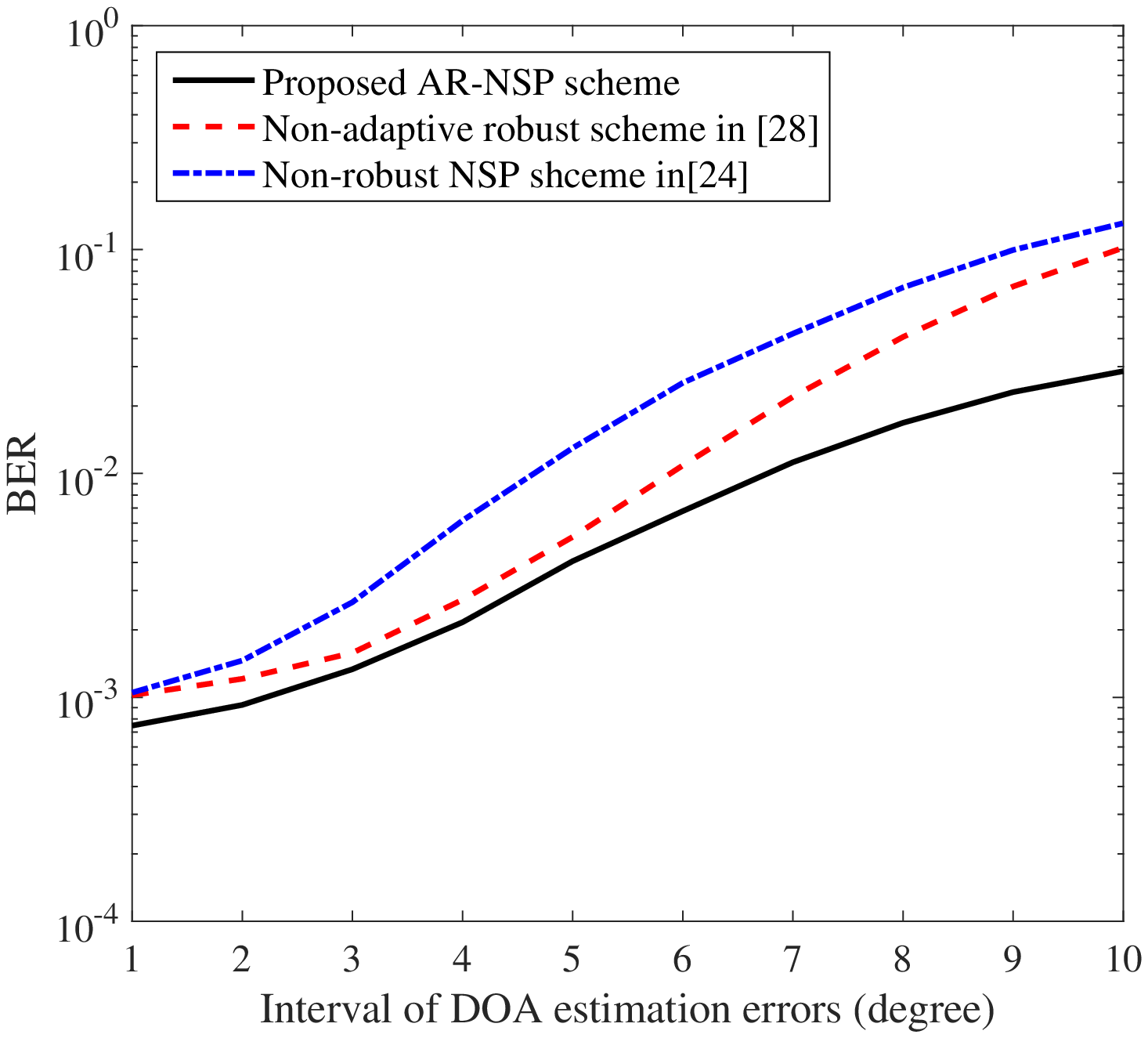}\\
  \caption{BER of three different schemes versus the value interval of DOA estimation errors (SNR=10dB, N=16).}\label{BER_MAX_ERROR_10dB}
\end{figure}

\begin{figure}[h]
  \centering
  \includegraphics[width=0.52\textwidth]{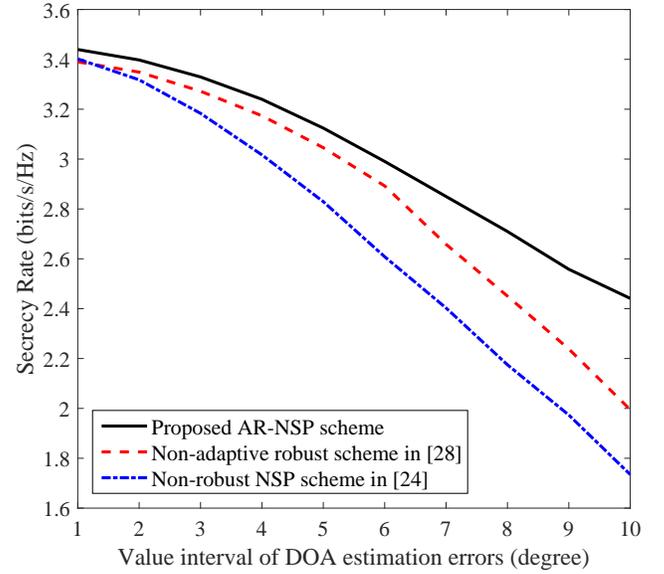}\\
  \caption{SR of three different schemes versus the value interval of DOA estimation errors (SNR=10dB, N=16).}\label{SR_MAX_ERROR_10dB}
\end{figure}

In Fig.~\ref{SNR_EST}, we plot the estimated SNR versus the exact SNR with different numbers of antennas at Alice.
In this figure, we first observe that the estimated SNR becomes closer to the exact SNR as the number of antennas increases. We also observe that the estimated SNR slightly deviates from the exact SNR when the exact SNR is less than 0dB and this deviation increases as the exact SNR decreases. However, as we can see from this figure, the estimated SNR is almost the same as the exact SNR when SNR is higher than 0dB. This means that the proposed SNR estimator can be viewed as an unbiased estimator when the exact SNR is relatively high.

In Fig.~\ref{RMSE_SNR}, we plot the root mean squared error (RMSE) of the SNR estimation versus the exact SNR with different values of $N$. In this figure, we first observe that the RMSE monotonously decreases with the exact SNR. In addition, we observe that as the number of antennas at Alice (i.e., the value of $N$) increases, the RMSE decreases. These two observations indicate that the estimation accuracy of the SNR can be improved by increasing the number of receive antennas at Alice or increasing the transmit power.

\begin{figure}[tp]
\centering
\subfigure[Phase versus direction angle of non-robust NSP scheme]{
\label{OP}
\includegraphics[width=0.49\textwidth]{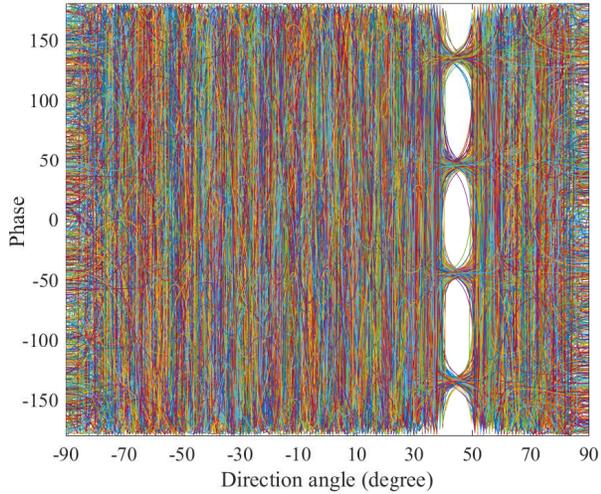}}
\subfigure[Phase versus direction angle of non-adaptive robust scheme]{
\label{MMSE}
\includegraphics[width=0.49\textwidth]{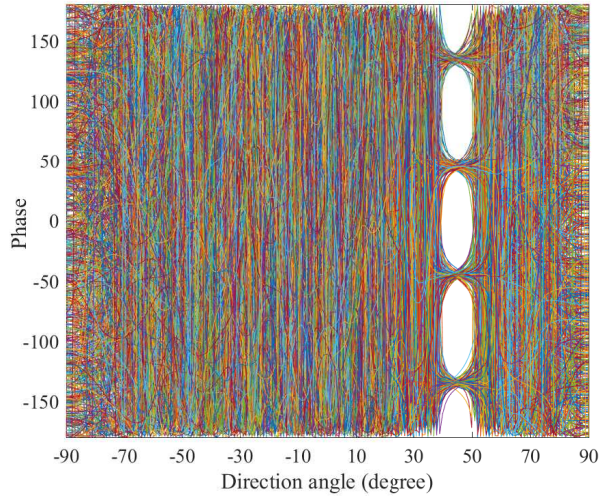}}
\subfigure[Phase versus direction angle of proposed AR-NSP scheme]{
\label{Proposed}
\includegraphics[width=0.49\textwidth]{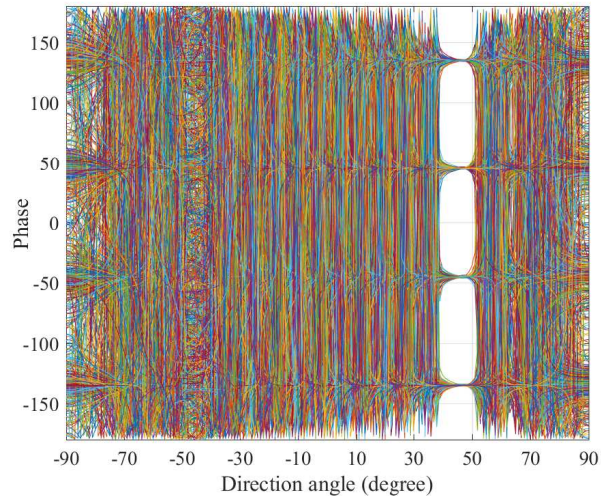}}
\caption{Eye patterns of three different schemes when the value interval of DOA estimation errors is 4 degree.}
\label{phase}
\end{figure}

In order to examine the accuracy of the estimated angle achieved by Root-MUSIC, in Fig.~\ref{CRB_SNR} we plot the RMSE of the estimation of $\theta$ versus the exact SNR with different numbers of antennas at Alice, while the corresponding CRLB is also presented as a benchmark. In this figure, we first observe that as the exact SNR increases, both the RMSE and CRLB monotonously decrease and the RMSE becomes closer to the CRLB. This demonstrates that the performance of ROOT MUSIC DOA estimation is better in the high SNR regimes than in the low SNR regimes. In addition, in this figure we observe that both the RMSE and CRLB decreases as the number of antennas at Alice increases, which shows that the direction angle estimation is more accurate when the numbers of antennas becomes larger. Furthermore, from this figure we can see that the RMSE is almost the same as the CRLB when the exact SNR is higher than 0dB. This means that the Root-MUSIC is a consistent estimator of direction angle. Finally, in this figure we observe that the RMSE converges to the CRLB more rapidly as the exact SNR increases when the number of antennas at Alice is larger.

In Fig.~\ref{rho_BER}, we plot the BER (i.e., bit error rate) achieved by the proposed AR-NSP scheme versus different values of $\rho$ with different SNRs in order to examine the impact of different values of $\rho$ on the secrecy performance of the proposed AR-NSP scheme. In this figure, we first observe that the BER of the proposed scheme is indeed a function of $\rho$ and there is an optimal value of $\rho$ that minimizes the corresponding BER. Surprisingly, under the specific system settings we observe that $\rho=1$ achieves the minimum BER, which is the reason that we adopt $\rho = 1$ in the proposed AR-NSP scheme. In addition, we observe that the BER is more sensitive to the values of $\rho$ when $\rho>1$ relative to when $\rho \leq 1$. As such, without enough information or resource to determining the optimal value of $\rho$, it is reasonable to set $\rho \leq 1$ in order to reduce the impact of $\rho$ on the secrecy performance of the proposed AR-NSP scheme.

In Fig.~\ref{BER_MAX_ERROR_10dB}, we examine the BER  performance of the proposed AR-NSP scheme, with the non-adaptive robust scheme proposed in \cite{Hu2016} and the non-robust NSP scheme proposed in \cite{Ding2014} as benchmarks. Specifically, in Fig.~\ref{BER_MAX_ERROR_10dB} we plot the BER of the three schemes versus the value interval of DOA estimation errors with SNR=10dB and N=16. In this figure, we observe that the BER of the proposed scheme is much lower than that of the other benchmark schemes. In general, the BER performance gap becomes more dominant as the value interval of DOA estimation errors increases. This demonstrates the superiority of the proposed scheme over the other two schemes. Furthermore, as expected we observe that the BER of the three schemes increases with the value interval of DOA estimation errors.

In Fig.~\ref{SR_MAX_ERROR_10dB}, we show the SR (i.e., secrecy rate) of the proposed AR-NSP scheme versus the value interval of DOA estimation errors, again with the non-adaptive robust scheme proposed in \cite{Hu2016} and the non-robust NSP scheme proposed in \cite{Ding2014} as benchmarks. In this figure, we first observe that the SRs of the three schemes decrease as the interval of DOA estimation errors increases. This is due to the fact that the probability of the estimated direction angle deviating from the exact direction angle becomes larger as the interval of DOA estimation errors becomes larger. In addition, we observe that the proposed AR-NSP scheme outperforms the other two schemes in terms of achieving a much higher SR. Furthermore, in this figure we observe that the SR performance gap between the proposed scheme and the other two schemes increases as the interval of DOA estimation errors increases. This can be explained by the fact that the beamforming vector of confidential information and projection vector of AN in the proposed AR-NSP scheme are averaged over different channel conditions, while they are set without considering different channel conditions in the other two schemes.

In order to examine the phase of signal constellation versus direction angle (namely eye patten) of the proposed AR-NSP scheme. Fig.~\ref{phase} presents the eye patterns of 1000 random QPSK symbols of the proposed AR-NSP scheme, the non-adaptive robust scheme proposed in \cite{Hu2016}, and the non-robust NSP scheme proposed in \cite{Ding2014}, where the interval of DOA estimation errors is 4 degree. From the three subfigures, we can see that the eye pattern of the proposed scheme in subfigure (c) opens larger and cleaner than those of the other two schemes presented in subfigure (a) and (b).  This means that the proposed method can reduce the effect of AN along the desired direction and push more AN towards the eavesdropper direction. This is the main reason that our proposed AR-NSP scheme can significantly outperform the other two schemes.
%
%

\section{Conclusion}
In this paper, we proposed the AR-NSP scheme to enhance PLS of DM systems in wireless networks. In the proposed scheme, the DOAs and SNRs of the desired user and eavesdropper are firstly estimated, based on which the value intervals of DOA estimation errors are predicted. Then, the beamforming vector of the confidential information and the AN projection vector are designed as per these estimated DOAs/SNRs and the predicted error intervals. Our examination shows that the proposed AR-NSP scheme can significantly outperform two benchmark schemes in terms of achieving a lower BER and a higher SR. In the context of DM systems, the proposed scheme eliminates the impractical assumption of applying PLS, which is that the exact or estimated locations of the desired user and eavesdropper are \emph{a priori} available. As such, the proposed scheme can potentially achieve the ever-lasting information-theoretic security in some practical scenarios, such as IoT,  future mobile communications, satellite communications, device-to-device, vehicle-to-vehicle, millimeter-wave communications, smart transportation, and unmanned-aerial-vehicles networks.

\section*{Acknowledgments}

This work was supported in part by the National Natural Science Foundation of China (Nos. 61771244, 61501238, 61702258, 61472190, and 61271230), in part by the Open Research Fund of National Key Laboratory of Electromagnetic Environment, China Research Institute of Radiowave Propagation (No. 201500013), in part by the Jiangsu Provincial Science Foundation under Project BK20150786, in part by the Specially Appointed Professor Program in Jiangsu Province, 2015, in part by the Fundamental Research Funds for the Central Universities under Grant 30916011205, and in part by the open research fund of National Mobile Communications Research Laboratory, Southeast University, China (Nos. 2017D04 and 2013D02).

\ifCLASSOPTIONcaptionsoff
  \newpage
\fi

\bibliographystyle{IEEEtran}
\bibliography{IEEEfull,cite}

\end{document}